\documentclass[preprint,12pt]{elsarticle}

\usepackage{amsmath}
\usepackage{amssymb}
\usepackage{graphicx}% Include figure files
\usepackage{dcolumn}% Align table columns on decimal point
\usepackage{bm}% bold math
\usepackage{yfonts}          % for the Gothic font used for the Q factor

\journal{Elsevier}

\begin{document}

\begin{frontmatter}



\title{Fluctuation-dissipation theorems for multi-phase flow with memory in porous media}


\author[inst1]{Dick Bedeaux}
\author[inst1]{Signe Kjelstrup}
\ead{signe.kjelstrup@ntnu.no}

\affiliation[inst1]{organization={PoreLab, Department of Chemistry, Norwegian University of Science and Technology, NTNU},%Department and Organization
        %    addressline={Address 1}, 
            city={Trondheim},
            postcode={7491}, 
         %   state={State One},
            country={Norway}}

\author[inst2]{Steffen Berg}
\ead{steffen.berg@shell.com}

\affiliation[inst2]{organization={Shell Global Solutions International B.V.},%Department and Organization
            addressline={Grasweg 31}, 
            city={Amsterdam},
            postcode={1031 WG}, 
         %   state={State Two},
            country={The Netherlands}}

\author[inst3]{Umar Alfazazi}
\author[inst3]{Ryan T. Armstrong}
\ead{ryan.armstrong@unsw.edu.au}

\affiliation[inst3]{organization={School of Civil and Environmental Engineering, The University of New South Wales},%Department and Organization
          %  addressline={Address Two}, 
            city={Syndey},
         %   postcode={22222}, 
            state={NSD},
            country={Australia}}

\begin{abstract}
Recent works have reported on the collective behavior of multiphase systems under fractional flow. Such behavior has been linked to pressure and/or flux fluctuations under stationary flow conditions that occur over a broad range of resonance frequencies and associated relaxation times. However, there currently exists no theoretical development to deal with such phenomena. The aim of this paper is to develop a fundamental theory that can describe such behavior. Fluctuation-dissipation theorems for the case with memory are formulated, providing a new route to obtain frequency-dependent porous media permeability. 

We propose that multiphase flow systems can be explained by a multipeak Lorentzian memory function and provide supporting experimental data from the flow of decane and water in a porous medium made of glass beads. Our fluctuation dissipation theorems provide information on different types of relaxation phenomena and resonance frequencies that occur during fractional flow. We show, using experimental data, that Green-Kubo-like expressions can be formulated for two-phase fluid flow driven by a constant pressure drop. The resulting autocorrelation functions, or rather their Fourier transforms, exhibit multiple Lorentzian peak shapes. Resonances are similar to those of electric conductance. 
The analysis offers a new route to steady-state relative permeability measurements, including information on the relaxation times and resonance regimes that exist during fractional flow. Overall, the theory presented and supported by fractional flow experiments provides a rich set of possible directions for future developments that could fundamentally change the way multiphase flow systems are understood and studied. 
\end{abstract}

\begin{graphicalabstract}
\includegraphics[scale=0.5]{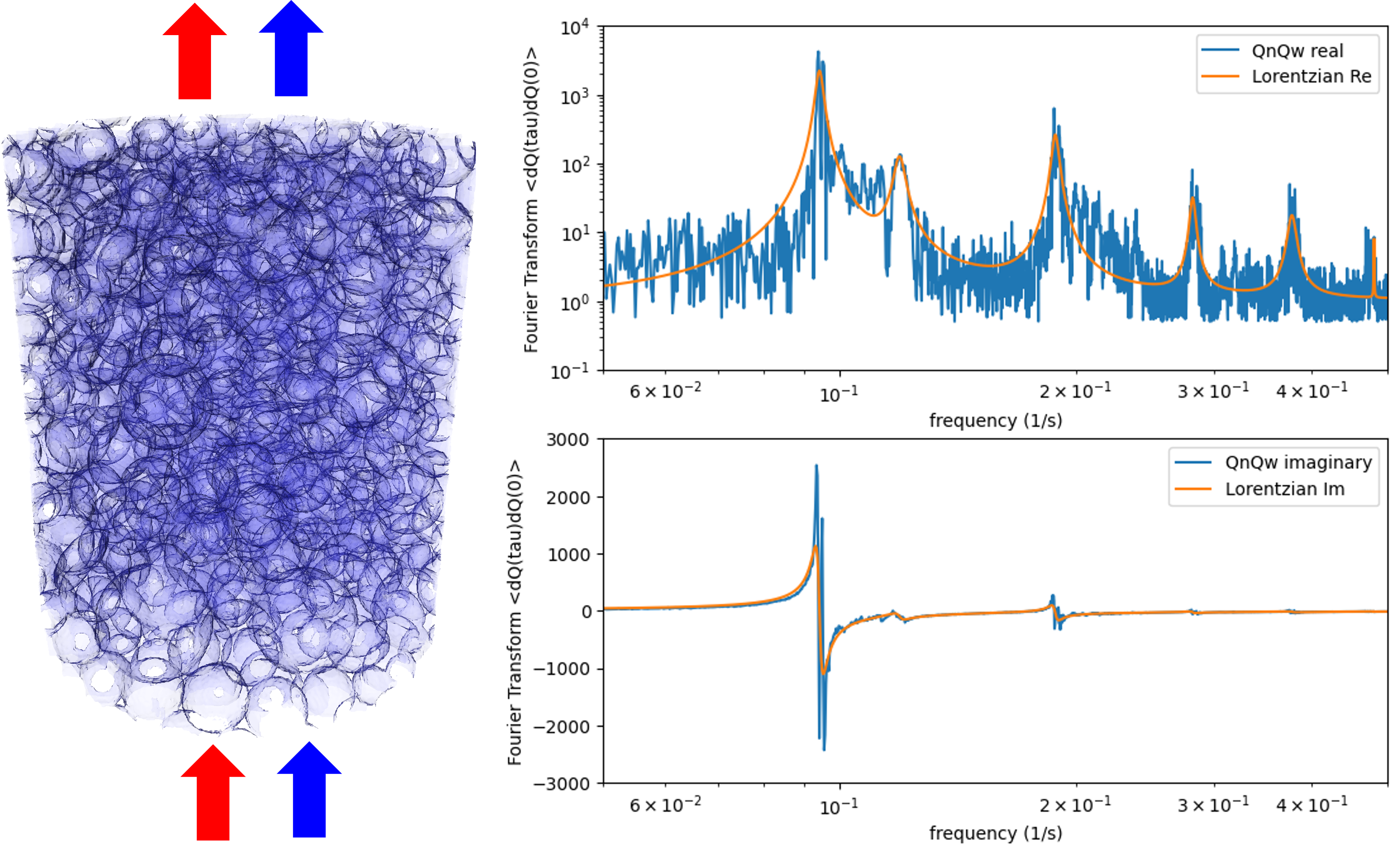}
\end{graphicalabstract}

\begin{highlights}
\item We formulate for the first time the fluctuation dissipation theorem (FDT) for the time and frequency - dependent permeability of a porous medium
\item We apply FDT to experimental data and obtain the frequency dependent permeability of two phase flow of decane and water in a mixture of glass beads of mm size.
\item In the linear force-flux regime the frequency spectrum of the flux autocorrelation exhibits a series of peaks mainly in the frequency range between 0.01 and 0.1 Hz 
\item The peak are well described by a Lorentzian with very small bandwidth implying a resonance phenomenon with high quality factor comparable to a tuning fork 
\item This points to cooperative processes such as ganglion dynamics as dominant transport process.
\end{highlights}


\end{frontmatter}


%\linenumbers


%=======================================
% Introduction
%=======================================

\section{Introduction}
\label{sect:introduction}

%\subsection{How does memory appear in the %context of porous media permeability}

Porous media present us with a vast class of transport systems, for instance, in biology, geology, or technology. The mechanisms of transport vary greatly with the pore size, adsorption to the walls, wettability, or magnitude of the driving force. Therefore, predictions of fluid permeability in porous media are of key interest. Permeability is experimentally determined using Darcy's law and its extension for multiphase flow~\cite{Whitaker1986B}. The experimentally observed fluctuations~\cite{DiCarlo2003} are commonly assumed to average out, yet current work has characterized them as nonthermal fluctuations related to intrinsic phenomena (and not instrumental noise) at highly different relaxation times \cite{Armstrong2013,Armstrong2014,Rucker2015,Schluter2017,Rucker2021,Spurin2023DMD}.

\begin{figure*}[ht]
\centerline{\includegraphics[scale=0.35]{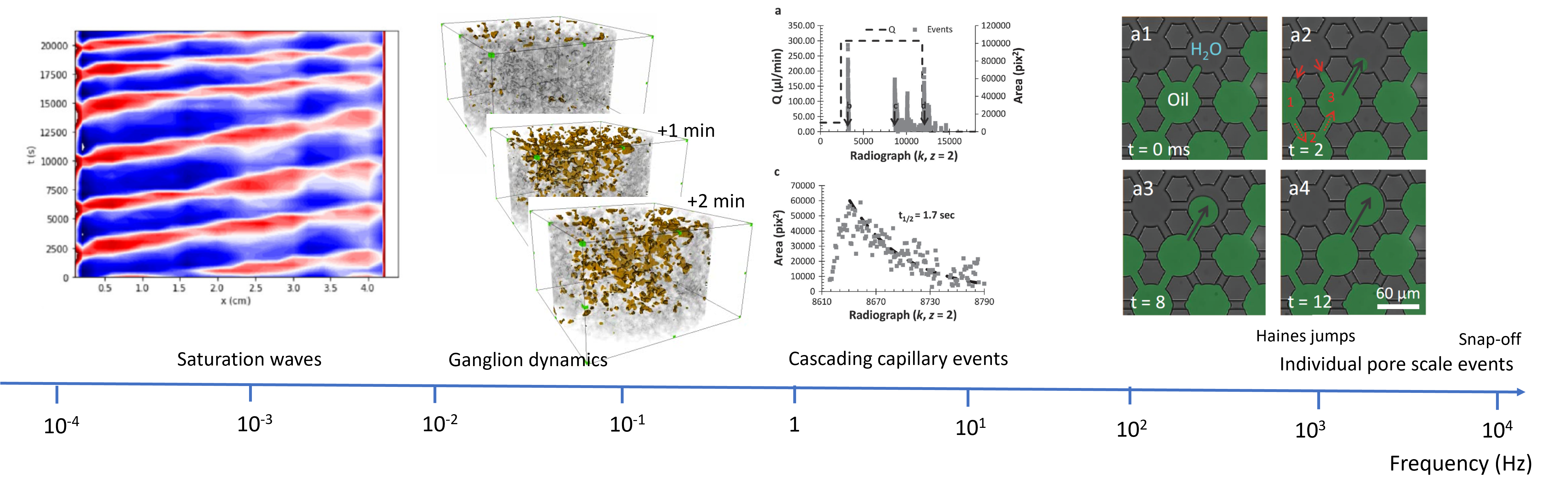}}
	\caption{Possible relaxing phenomena in multi-phase flow in porous media within  typical frequency ranges~\cite{Rucker2021,Rucker2015,Armstrong2014,Armstrong2013}.}
	\label{fig:frequencyoverview}
\end{figure*}

Figure~\ref{fig:frequencyoverview} presents an overview of typical phenomena in porous media fluid transport. The traveling saturation waves that occur on a time scale of minutes \cite{Rucker2021} have the longest relaxation times (shown on the left side). They are followed by ganglion dynamics \cite{Rucker2015} and cascading capillary events on the second to some fractions of a second scale time scale \cite{Armstrong2014}. The dynamics of the interface of individual pore-scale events, such as Haines jumps~\cite{Haines1930}, can take place on the millisecond time scale \cite{Armstrong2013} and are illustrated on the right-hand side. 

The frequency ranges corresponding to these phenomena and their respective relaxation times are shown at the bottom of the figure. The behavior of a system changes depending on whether the system is driven slower or faster than the relaxation time. For example, polymer solutions behave viscous when driven slower than the relaxation time and elastic when driven faster than the relaxation time, and when flowing in porous media, the flow regimes change, respectively \cite{Ekanem2020}. Another example is the dielectric constant that changes depending on whether the frequency lies below or above a relaxation phenomenon related to the charge distribution~\cite{Kremer2002,Liu2012}, which is then associated with memory effects. Therefore, we expect that the behavior of multiphase flow is a function of the relaxation times associated with the phenomena in Fig.~\ref{fig:frequencyoverview}. In turn, it can be expected that the time signal contains information on the mechanism of transport. 

Although the analysis of relaxation times is very common in other research domains such as in the dielectric behavior of materials~\cite{Kremer2002,Liu2012}, as of now there is no systematic procedure in the science of porous media, which deals with the memory effects in flux correlations, in order to extract information on porous media permeabilities. Correlations, per se, imply memory effects.  We have recently proposed to use noise correlation functions from steady-state experiments and have reported a procedure to find the zero frequency permeability of two-phase flow in a porous medium \cite{Alfazazi2024}. 

The fluctuation dissipation theorem (FDT) deals with time correlation functions. The zero frequency permeability was obtained from the integral over time and therefore contains a memory function. In their first report Alfazazi \textit{et al.} \cite{Alfazazi2024} found that the volume autocorrelation function was related to Darcy law permeability by a factor $F$. The volume flow autocorrelation function contained memory effects, but these were not further examined in the first study. A study of these is therefore the topic of the present paper. 

%\subsection{The memory function - a new path to transport coefficients}

A full analysis of the permeability memory function is of interest, and our central focus herein. We shall see that the analysis may offer a new route to steady-state permeability measurements, including their relaxation times and resonance regimes. The aim of this work is to improve the fundamental understanding of the transport properties of porous media by adding a method of analysis of memory signals. Transport in porous media is described on a macroscopic scale (continuum), and the construction of average properties is central. It has long been a challenge to find good bottom-up descriptions of the complex heterogeneous structure of the medium and flow patterns. The challenge remains when we want to apply FDT and deal with memory effects.

%ethods that describe fluids, confined in nanopores, are also scarce in spite of the documented large impact of confinement \cite{Braun2015}. 
%=============================
%\subsection{State-of-the-art}
%=============================
%\subsection{Modelling multiphase flow}

Over the past several decades, several attempts have been made to upscale pore-scale transport to the Darcy scale. A range of methods have been explored, such as volume-averaging methods, or using, e.g. Stokes or Navier-Stokes flow at the pore scale as a starting point. Other approaches have as a starting point the mass, momentum and energy balances, but are still based on volume averaging \cite{Whitaker1986A,Whitaker1986B}. Roy and coauthors \cite{Roy2022} pioneered a method using non-thermodynamic variables that obey the Euler homogeneous function. Our approach differs by linking fluxes and driving forces at the REV level following the general framework of nonequilibrium thermodynamics \cite{Kjelstrup2018,Kjelstrup2019,Bedeaux2023}. 

The way in which the variables are averaged over the representative elementary volume (REV) is crucial for the description \cite{Bear1972,Bedford1983,Nordahl2008}. In a recent definition, the REV was constructed to contain all micro-states of the medium \cite{Bedeaux2022,Bedeaux2023}. The focus is on the extensive REV variables. Geometric variables are included, following Hill \cite{Hill1994}. These can be used to handle confinement. The central building block is the REV, for which we define the average extensive variables (see \cite{Bedeaux2022,Bedeaux2023}). We shall describe transport and fluctuation for fluxes across a REV, \textit{i.e.}, the correlations of REV variables and their memory effects. Regardless of the approach, it has been shown that the choice of variables on the continuum scale is central \cite{McClure2018}. If central state variables are missed, the macroscale behavior may appear hysteretic~\cite{McClure2018}. Real hysteresis can occur due to physical processes or memory effects \cite{Gharari2018,Holtzman2020}. 

Hydrodynamic fluctuations are well documented for bulk fluids \cite{Ortiz2006}. Transport processes in porous media occur on a wider range of time scales. Joekar-Niasar \textit{et al.} observed relaxation behavior in dynamic capillary pressure curves~\cite{Niasar2010}. Winkler \textit{et al.} \cite{Winkler2020} observed athermal relaxation times for steady-state two-phase flow ranging from 10$^{-3}$ to 0.1 seconds for viscous two-phase flow in a network. When including coalescence dynamics, interface phenomena, and pore scale dynamics, the scale was extended to both ends, down to 10$^{-5}$ and up to 1-10 seconds \cite{McClure2021}. Changes in fluid configuration take from seconds \cite{Armstrong2014,Schluter2017,Rucker2021,Spurin2022} to
minutes \cite{Gao2021} and hours \cite{Schluter2017}, cf. Fig.~\ref{fig:frequencyoverview}. 
%
%\subsection{Which method and system do we use and why} 
%
In order to address the very wide range of physically relaxing phenomena and driving forces present in porous media, it is impractical to start the upscaling process at the pore-scale equation level, e.g. Stokes flow equation. Coupling might then be difficult or impossible to address. Several challenges must be encountered, such as complex closure relations \cite{Whitaker1986A,Whitaker1986B,Lasseux2022}. In addition, central-state variables might be missed. Therefore, the starting point pursued here is a general thermodynamic framework that links fluxes and driving forces at the REV level, with state variables included in a systematic manner \cite{Bedeaux2022,McClure2022}. 

A suitable thermodynamic theory of transport must be
able to describe the interplay or coupling of several forces as well as the energy dissipation involved (the entropy production). The theory of non-equilibrium thermodynamics (NET) \cite{deGroot1984,Kjelstrup2020} provides such a basis, also when extended to porous media
\cite{Kjelstrup2018,Kjelstrup2019,Bedeaux2022,Bedeaux2023}. The extension rests on the assumption of the REV construction; that a REV exists with average variables defined so that all of the microstates are captured and the Gibbs equation is obeyed for the (average) variables. Using this theory, we shall exploit the characteristic signature of the fluctuations in each flux of the REV. We are thus making the bold assumption that FDT applies to the set of coarse-grained variables that define our REV. The long-range aim is to be able to find permeability properties of the porous media on this basis, exploiting the memory of the available signals. 

%\subsection{The aim of this investigation} 
%
Fluctuation-dissipation theorems \cite{Kubo1966} have been used for a long time to provide information on transport properties. In practice, there is a shortest accessible time scale for an experiment of interest, as well as for a numerical simulation. This implies that the integral over the correlations applied in FDT~\cite{Callen1951,Green1954,Kubo1966} may be missing out on fast decaying contributions. We will investigate how this affects FDT. 
The aim of this work is to elucidate conditions for these effects and other issues related to the memory functions connected to the permeability. In this manner, we hope to make a new step in the direction of describing the permeabilities of porous media.

In the following theory section, we derive the equations for the time- and frequency-dependent variables of the REV. For details of the definition of REV variables and the derivation of entropy production relevant to multiphase fluid flow in porous media, we refer to Appendix A and previous work \cite{Kjelstrup2018,Kjelstrup2019,Bedeaux2022,Bedeaux2023}. The memory function is first presented in Section \ref{sec:mem}. In Section~\ref{sect:constituvivefrequency} we discuss frequency-dependent constitutive equations. Section~\ref{sect:multiresonance} presents a Lorentzian multiple resonance model, which is later compared with our experimental results. In
Section~\ref{sect:fluctuationdissipation} we add the fluctuating contributions to the fluxes and provide the corresponding FDT.  We then discuss how the theorems provide direct routes to the frequency-dependent Onsager coefficient \cite{Onsager1931A,Onsager1931B} of the system and therefore to the typical two-phase flow permeability. Together, we present a new method to determine
frequency-dependent transport coefficients in Section~\ref{sect:experimental}. In Section~\ref{sect:results} we discuss the results for the chosen example and how the method can offer insight into the flow of fluids in porous media.

\section{Theory}
A key characteristic of systems with memory is that the thermodynamic forces and fluxes are time dependent. Therefore, the flux-force relationships are time dependent as well. We start by defining a memory function that represents the history-dependent effects in the response of a system to external perturbations or fluctuations. The memory function is related to the system's memory or relaxation timescales. We then transform the flux-force relation with memory from the time domain to the frequency domain. This allows us to represent the memory function as a frequency-dependent parameter, well aligned with the common Lorentzian resonance model. A multiple resonance model is developed as a memory function that captures the various frequencies and relaxation times associated with multiphase flow. Lastly, we propose that the statistical properties of the fluctuating fluxes, as captured by their ensemble average, are related to the temporal correlation structure of the system, as described by the FDT and related memory function. This last part provides a way to relate experimentally measured fluctuations to continuum-scale transport properties, such as the permeability used by Darcy's law. 

%=======================================
\subsection{The memory function}
%=======================================
\label{sec:mem}

The entropy production was written for a REV in a porous medium \cite{Kjelstrup2018,Kjelstrup2019,Bedeaux2022,Bedeaux2023}. The derivation is outlined in Appendix A for the two immiscible and isothermal fluid phases. There are two flux-force products: 
\begin{equation}
\sigma (t) =-\frac{1}{T}J_{w}(t)\frac{\partial }{\partial x}\mu _{w}(t) -\frac{1}{T}%
J_{n}(t)\frac{\partial }{\partial x}\mu _{n}(t)  \label{1.6}
\end{equation}%
where $J_w$ and $J_n$ are fluxes of the more wetting ($w-$)phase and the less wetting ($n-$) phase, $\mu_w, \mu_n$ are the corresponding chemical potentials, and  $T$ is the temperature. All variables are functions of time, and the transport is unidirectional.  The
thermodynamic forces under isothermal conditions are%

\begin{equation}
-\frac{1}{T}\frac{\partial \mu _{i}(t)}{\partial x} = - \frac{V_i}{T} \frac{\partial p (t)}{\partial x} \label{2.5}
\end{equation}%
The dimension of the force is J.K$^{-1}$.m$^{-1}$.mol$^{-1}$. The driving forces are introduced in the entropy production, and we obtain:

\begin{equation}
\sigma (t) =-\frac{1}{T} \left [ V_wJ_{w}(t) + V_nJ_n(t) \right ]  \frac{\partial p (t)}{\partial x}  = -J_V(t) \frac{1}{T} \frac{\partial p (t)}{\partial x} 
\label{1.7}
\end{equation}
The volume flux $J_V(t)$ is
\begin{equation}
J_{V}(t)= V_{w}J_{w}(t)+V_{n}J_{n}(t)
\label{2.8}
\end{equation}
The linear flux-force relation with memory has thus only one flux and one driving force:
\begin{equation}
  J_{V}(t) =  -\frac{1}{T}\int_{t
}^{\infty }L_{VV}(t-t^{\prime })\frac{\partial p(t^{\prime })}{\partial x}%
dt^{\prime } 
\end{equation}
The transport coefficient $L_{VV}(t-t^{\prime })$ describes the memory of the past and is therefore called the memory function for short. The memory function is zero for negative arguments because of causality. The formula gives the time-dependent permeability in a system with an oscillating driving force. 
The volume flux has dimension m.s$^{-1}$, while the hydraulic permeability $L_{VV}$  has dimension m$^5.$K.s$^{-1}$.J$^{-1}$. 

The permeability can be a function of state variables, but not of the driving force or the flux. The state variables are here properties of the REV. A change in geometry of the porous medium may change the state variables. A change in flow boundary conditions may lead to changes in the state of the REV and therefore change the permeability.  
%=================================================
% Frequency dependent constitutive equations
%=================================================

\subsection{Frequency dependent constitutive equations}\label{sect:constituvivefrequency}
Memory effects are more conveniently described by Fourier
transforms. For the frequency-dependent volume flux, we have: 
\begin{equation}
J_{V}(\omega )\equiv \int_{-\infty }^{\infty }e^{i\omega t}J_{V}(t)dt
\label{3.1}
\end{equation}%
while the frequency-dependent pressure gradient is:
\begin{equation}
\frac{\partial p (\omega)}{\partial x} \equiv \int_{-\infty }^{\infty }e^{i\omega t}  \frac{\partial p (t)}{\partial x}  dt 
\label{3.2}
\end{equation}%
 The dimension of the flux is now in m. 
When we Fourier-transform the linear flux-force relation (5), we obtain
\begin{equation}
J_{V}(\omega )=\frac{1}{T}L_{VV}(\omega )\frac{\partial p(\omega )}{\partial
x}  \label{3.4}
\end{equation}
The force, flux, and the memory function are all real functions of time. 
This implies for the frequency dependent memory function that%
\begin{equation}
L_{VV}^{\ast }(\omega )=L_{VV}(-\omega )  \label{3.5}
\end{equation}%
Superscript $^{\ast }$ means complex conjugate. When $L^{\ast}_{VV}(\omega_0) = L_{VV}(-\omega_0)$ near a resonance are large, the flux resulting from the oscillating driving force becomes large.

Causality means that%
\begin{equation}
L_{VV}(\tau )=0\rm{ \ \ for \ \ }\tau <0  \label{3.6}
\end{equation}%
This implies that the real and the imaginary part of the fequency-dependent permeability, $L_{VV}(\omega )= L_{VV}^{\prime }(\omega )+iL_{VV}^{\prime \prime }(\omega )$, satisfy the Kramers-Kronig relations;  .
\begin{eqnarray}
L_{VV}^{\prime }(\omega ) &=&\frac{1}{\pi }\mathcal{P}\int_{-\infty
}^{\infty }\frac{L_{VV}^{\prime \prime }(\omega ^{\prime })}{\omega
^{\prime }-\omega }d\omega ^{\prime }  \nonumber \\
L_{VV}^{\prime \prime }(\omega ) &=&-\frac{1}{\pi }\mathcal{P}%
\int_{-\infty }^{\infty }\frac{L_{VV}^{\prime }(\omega ^{\prime })}{%
\omega ^{\prime }-\omega }d\omega ^{\prime }  \label{3.7}
\end{eqnarray}%
Here $\mathcal{P}$\ is the principal value. 

Consider next a force oscillating with the frequency $\omega _{0}$. We define for simplicity:%
\begin{equation}
X_{V}(t)\equiv - \frac{1}{T} \frac{\partial p(t)}{\partial x} = X_{V,0}\cos (\omega _{0}t)  \label{3.10}
\end{equation}
The Fourier transform is 
\begin{equation}
X_{V}(\omega )=\pi X_{V,0}\left[ \delta (\omega -\omega _{0})+\delta (\omega
+\omega _{0})\right]   \label{3.11}
\end{equation}%
where $ X_{V,0}$ is the amplitude of the oscillating force. 
The resulting volume flux is%
\begin{eqnarray}
\begin{aligned}
J_{V}(\omega ) &=& L_{VV}(\omega) X_{V}(\omega )   \\
&=&\pi X_{V,0} [ L_{VV}(\omega_0) \delta (\omega -\omega _{0})  \\ 
& & + L_{VV}(-\omega _{0}) \delta (\omega +\omega _{0}) ] \label{3.12}
\end{aligned}
\end{eqnarray}%
Fourier transforming this back in time gives%
\begin{eqnarray}
J_{V}(t) &=&\frac{1}{2} X_{V,0} [L_{VV}(\omega _{0}) e^{-i\omega _{0}t}  +L_{VV}(-\omega
_{0}) e^{i\omega _{0}t} ] \notag
\\
\label{3.13}
\end{eqnarray}%
 The
entropy production becomes %
\begin{eqnarray}
\begin{aligned}
\sigma (t) &=&J_{V}(t) X_{V}(t)   \\
&=& X_{V,0}[L_{VV}^{\prime }(\omega _{0})( 1 + \cos 2\omega_{0}t )  \\
& & -L_{VV}^{\prime \prime }(\omega _{0})(\sin 2\omega _{0}t)]X_{V,0}
\label{3.14}
\end{aligned}
\end{eqnarray}%
 By averaging the outcome over half a period, we obtain%
\begin{equation}
\bar{\sigma} = L_{VV}^{\prime }(\omega _{0})X_{V,0}^2
\label{3.15}
\end{equation}%
The cosine or sine dependency drops out in this expression. To satisfy the second law, the matrix $L_{VV}^{\prime }(\omega
_{0})$\ has to be positive semidefinite. 

In terms of the pressure gradient and the memory function of the volume flow, the average entropy production is equal to
\begin{equation}
\bar{\sigma} =L_{VV}^{\prime }(\omega _{0})\left[ \frac{1}{T}\left( \frac{\partial
p}{\partial x}\right) _{0}\right] ^{2}  \label{3.15b}
\end{equation}
 The average entropy production is constant and does not depend on memory effects, as expected. The subscript 0 indicates the force amplitude.
%=================================================
% Multiple resonance model
%=================================================

\subsection{A Lorentzian multiple resonance model}
\label{sect:multiresonance}
As will become clear from the experiments that follow, Lorentzian memory functions are relevant.  We include $m$ resonances and find satisfactory fits. The behavior can be compared with that of conductive materials \cite{Kremer2002,Liu2012}.

The data are interpreted in terms of frequencies rather than angular frequencies. Their interrelation is $f = \omega/ (2 \pi)$.
The frequency-dependent memory function is then: 
\begin{equation}
L_{VV}(f )= \sum_{i=1}^{m} L _{VV}^{i,0} 
\left[ \frac{2\Gamma^i f} {
2 \Gamma ^i f - i ((f ^{i})^2  - f^{2})}  \right]   \label{3.17}
\end{equation}%
where $\Gamma^i$ is the bandwidth, $\tau ^{i} = 1/\Gamma^i$ is the relaxation time, and $f^{i}$ is the resonance frequency of the contribution $i$. 
Given that all relaxation times are positive, all poles are in the lower half of the complex $f$- plane. This property is consistent with the
fact that $L _{VV}(t)=0$ for $t<0$ and is therefore a consequence of
causality. The poles are in pairs. The real part of $L_{VV}(f )$ 
is then%
\begin{equation}
\text{Re} L _{VV}(f ) =
 \sum_{i=1}^{m} L _{VV}^{i,0} 
 \frac{(2\Gamma^i f )^2}{ ((f^i)^2- f^2)^2 + (2 \Gamma^i f)^2}
\label{3.19}
\end{equation}%
and the imaginary part is%
\begin{eqnarray}
\text{Im}L_{VV}(f ) &=&  \sum_{i=1}^{m} L_{VV}^{i,0} 
 \frac{2\Gamma^i f ( (f^i)^2- f^2) }{ ((f^i)^2- f^2)^2 + (2 \Gamma^i f)^2}  \label{3.20}
\end{eqnarray}%

%===========================
When fitting the experimental data, we shall use both real and imaginary parts. The approximation $f ^{i} \tau
^{i}>>1$ turns out to be valid in the present case, facilitating the determination of the amplitude $L_{VV}^{i,0}$  of each resonance $i$. 

In experiments and network simulations, there is usually a smallest time scale $\tau _{0}$. When we measure or calculate $L _{VV}(t)$,
contributions with $\tau< < \tau _{0}$ do not contribute \cite{Alfazazi2024}. For the Fourier transform, this implies that resonances with $f^i >> 1/\tau_0 $ will not contribute.

%=================================================
% Fluctuation-dissipation theorems
%=================================================

\subsection{Fluctuation-dissipation theorems}
\label{sect:fluctuationdissipation}

We described above how the two-phase flow through a porous medium is governed by the entropy production. The variables of the description were averages constructed for the REV, the description is applied on a course-grained scale, the REV scale, and not the pore scale \cite{Bedeaux2022,Bedeaux2023}. 

This coarse-grained description is then extended to incorporate fluctuations inside the REV. We extend the description by adding fluctuating contributions to the fluxes, contributions which are indicated by the subscript $R$. They are crucial in the formulation of fluctuation-dissipation theorems for homogeneous systems \cite{Callen1951,Green1954,Kubo1966} and will now be formulated for porous media. On a molecular scale, they
are changing rapidly and their spatial correlations
are on the molecular scale. On the REV scale (the coarse-grained scale)
the correlations in space refer to the macroscopic spatial scale. Their source strength is given by $2k_{\rm{B}}$ times the Onsager conductivity matrix \cite{Onsager1931A,Onsager1931B}. As we shall see, the
temporal correlation function is related to the memory function.  

For the volume flux, we obtain 
\begin{equation}
J_{V,tot}=J_{V}+J_{V,R}  \label{4.1}
\end{equation}%
The flux is a local-valued flux and not a flux which is averaged over the cross-sectional area of the porous medium (normal to the direction of flow). 
The average of the random contribution to the flux is equal to zero. 
\begin{equation}
\left\langle J_{V,R}\right\rangle =0  \label{4.3}
\end{equation}%
This implies that%
\begin{equation}
\left\langle J_{V,tot}\right\rangle =\left\langle J_{V}\right\rangle 
\label{4.4}
\end{equation}
The typical flow average with noise is presented in the experimental section. The second moments of the signal follow from the statistical mechanical description.  

We shall now make the bold assumption that the fluctuations satisfy the fluctuation-dissipation theorems \cite{Callen1951,Green1954,Kubo1966} for the REV. This means that:

\begin{equation}
\begin{aligned}
  \left\langle J_{V,R}(\mathbf{r},t)J_{V,R}(\mathbf{r}^{\prime },t^{\prime
})\right\rangle  &=&\left\langle J_{V,R}(\mathbf{r},t)J_{V,R}(\mathbf{r}
^{\prime },t^{\prime })\right\rangle \\
& = & 2k_{\rm{B}}L_{VV}(t-t^{\prime
})\delta (\mathbf{r}-\mathbf{r}^{\prime })  
 \label{4.5}  
\end{aligned}    
\end{equation}
Here we use the fact that the memory kernel as a function of time is
real. The $\delta $-functions reflect the short-range nature of the spatial
correlations. The REV is chosen so that we do not have a non-local description. The REV must include coarse-grained pore-scale events such as Haines jumps \cite{Haines1930} or specific space-time correlations
\cite{Armstrong2013,Armstrong2014,Bedeaux2022}. In this work, we do not consider fluctuations in the heat flux.

For the Fourier transforms of the fluxes, we find from Eq.\ref{4.5}:
\begin{equation}
\begin{aligned}
\left\langle J_{V,R}^{\ast }(\mathbf{r},\omega )J_{V,R}(\mathbf{r}^{\prime
},\omega ^{\prime })\right\rangle  &=&\left\langle J_{V,R}^{\ast }(\mathbf{r}
, \omega )J_{V,R}(\mathbf{r}^{\prime },\omega ^{\prime })\right\rangle \\
& = & \frac{%
k_{\rm{B}}}{\pi }L_{VV}(\omega)\delta (\omega +\omega ^{\prime
})\delta (\mathbf{r}-\mathbf{r}^{\prime })  
 \label{4.7}
\end{aligned}    
\end{equation}
where $\ast$ denotes a complex conjugate. 
In experiments and network simulations, there is usually a smallest time
scale $\tau _{0}$. This implies that contributions that decay on a faster
time scale are not seen, as we explained in the previous section. Equation %
\ref{4.5} should then be replaced by

\begin{equation}
\begin{aligned}
\left\langle J_{V,R}(\mathbf{r},t)J_{V,R}(\mathbf{r}^{\prime },t^{\prime
})\right\rangle  \\
= \left\langle J_{V,R}(\mathbf{r},t)J_{V,R}(\mathbf{r}^{\prime},t^{
\prime})\right\rangle \\
 =    2k_B \Theta (t-t' - \tau_0)L_{VV}
(t-t^{\prime })\delta (\mathbf{r}-\mathbf{r}^{\prime }) \\
= 2k_{\rm{B}}L_{VV}^{\rm{exp}
}(t-t^{\prime })\delta (\mathbf{r}-\mathbf{r}^{\prime }) 
  \label{4.8}
\end{aligned}    
\end{equation}

where $\Theta$ is the Heaviside function, which is one for a positive argument and zero for a negative argument. The equation defines $L_{VV}^{\rm{exp}}$. 
The Fourier transform becomes

%\begin{eqnarray}
%\left\langle J_{V,R}^{\ast }(\mathbf{r},\omega )J_{V,R}(\mathbf{r}^{\prime
%},\omega ^{\prime })\right\rangle  &=&\left\langle J_{V,R}^{\ast }(\mathbf{r}%
%,\omega )J_{V,R}(\mathbf{r}^{\prime },\omega ^{\prime })\right\rangle =\frac{%
%k_{\rm{B}}}{\pi }L _{VV}^{\rm{exp}}(\omega )\delta (\omega +\omega
%^{\prime })\delta (\mathbf{r}-\mathbf{r}^{\prime })  \nonumber \\
%&&  \label{4.9}
%\end{eqnarray}

\begin{equation}
\begin{aligned}
\left\langle J_{V,R}^{\ast }(\mathbf{r},\omega )J_{V,R}(\mathbf{r}^{\prime
},\omega ^{\prime })\right\rangle  &=&\left\langle J_{V,R}^{\ast }(\mathbf{r}
,\omega )J_{V,R}(\mathbf{r}^{\prime },\omega ^{\prime })\right\rangle \\
& = & \frac{%
k_{\rm{B}}}{\pi }L _{VV}^{\rm{exp}}(\omega )\delta (\omega +\omega
^{\prime })\delta (\mathbf{r}-\mathbf{r}^{\prime }) \\
\label{4.9}
\end{aligned}    
\end{equation}

In practice,  we have seen that the permeability from a Darcy law experiment, when compared to the result of the fluctuation-dissipation theorems in Eq.\ref{4.9}, deviate by a factor $L _{VV}^{\rm{exp}}(\omega )/L_{VV}(\omega )$, see \cite{Alfazazi2024} for the zero-frequency case. Therefore, the application of FDT is not straightforward.

We shall measure the average volume flow, $Q$, which is defined by:
\begin{equation}
    Q(x,t) \equiv \frac{1}{A} \int_A dy dz J_V(x,y,z,t)
    \label{eq:4.10}
\end{equation}
Here $A$ is the cross-sectional area of the REV perpendicular to the direction of transport.
Similar relations can be written for $Q_R$ in terms of $J_{V,R}$ and their Fourier transforms. 

For $Q_{R}$ Eqs.\ref{4.8} and \ref{4.9} give
\begin{equation}
    \left\langle Q_{R}(x,t)  Q_{R}(x',t') \right\rangle = \frac{2k_B}{A} L_{VV}^{\rm{exp}}(t-t') \delta (x-x^{\prime })
    \label{eq:4.13}
\end{equation}
where we have assumed that $L_{VV}^{\rm{exp}}$ is position-independent. Equation \ref{eq:4.13} becomes
\begin{equation}
    \left\langle Q^{\ast}_{R}(x,\omega)  Q_{R}(x',\omega') \right\rangle = \frac{k_B}{\pi A} L_{VV}^{\rm{exp}}(\omega) \delta (\omega+\omega') \delta (x-x^{\prime })
    \label{eq:4.14}
\end{equation}
The volume flow averaged over the $x-$coordinate, $Q(t)$, is defined by.
\begin{equation}
    Q_R(t) \equiv \frac{1}{L} \int_0^L dx Q_R(x,t)
    \label{eq:4.11}
\end{equation}
where $L$ is the length of the sample.  It follows from Eq.\ref{eq:4.13} that 
\begin{equation}
\left\langle Q_{R}(t)  Q_{R}(t') \right\rangle = \frac{k_B}{V_L} L^{\rm{exp}}_{VV}(t-t')  
\end{equation}
and 
\begin{equation}
\left \langle Q_{R}(\omega)  Q_{R}(\omega') \right \rangle = \frac{2k_B}{\pi V_L} L^{\rm{exp}}_{VV}(\omega)\delta (\omega + \omega')  
\label{eq:34}
\end{equation}
where $V_L = AL$. 

This theory concerns flux-fluctuation correlations. Fluctuations in pressure that are more accessible by experiment ~\cite{DiCarlo2003} must be dealt with otherwise.  

%=======================================
% Experimental system
%=======================================

\section{Experimental}\label{sect:experimental}

We refer to experimental procedures that have been described in detail elsewhere \cite{Alfazazi2024}. We repeat the essentials about the system and the measuring procedure so that the basis of the present results can be understood. A range of experiments were conducted under steady-state conditions. 
Two-phase flow through a porous glass bead pack was examined under a range of different pressure drops. We present experimental results from \cite{Alfazazi2024} that show memory effects and give a linear flux-force relationship. 

\begin{figure}[ht]	\centerline{\includegraphics[scale=0.5]{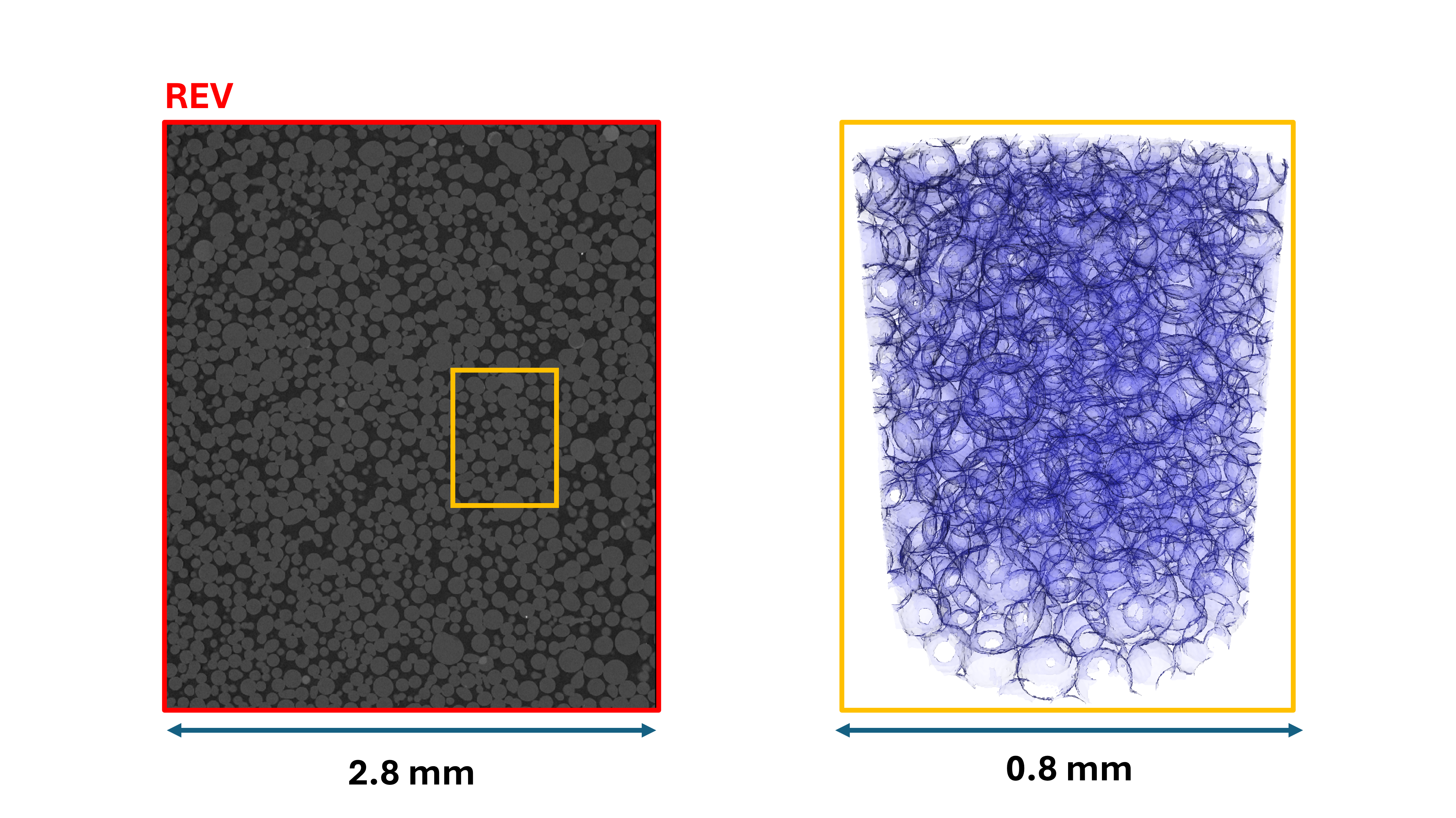}}
	\caption{Sintered glass beads used for two-phase flow experiments. The close-up to the right is a part of the representative elementary volume (REV) to the left.}
	\label{fig:Experimental_System}
\end{figure}

%\subsection{The porous medium}
%
A sintered glass sample (6 mm long and 3 mm diameter) was used for the coinjection experiment. The sample was generated by mixing two different groups of glass beads in the same proportion. One group of sizes was in the range of 0.09 to 0.15 mm. The other group of sizes was in the range of 0.1 to 0.2 mm. The sample was imaged with X-ray computed microtomography (2.7 micrometer resolution) and visualized in Figure \ref{fig:Experimental_System}. The mean pore size and porosity based on the segmented image (using Avizo software, ThermoFisher) were estimated at 60 micrometers and 31 percent, respectively. The absolute permeability in Darcy's law was determined to be approximately 10 Darcy. 
\noindent As immiscible fluids, we used DI-water and decane with an interfacial tension of approximately 30~mN/m. The densities of water and decane are $997$ kg/m$^3$ and $730$ kg/m$^3$, respectively.  Fluids are fully immiscible.  The glass beads are water-wet with a contact angle of approximately 30 degrees.

\begin{table}[ht]
\caption{Experiments used for analysis of memory effects}
	\centering
		\begin{tabular}{c|c|c|c|c} \hline
Exp. & Pressure drop & Capillary no & Flux & Variance \\
no.	 &	Psi	 & no. $\times 10^{-6}$ & $\times 10^{-5}$m/s & $ \times 10^{-12}$ \\ \hline
1 & 1.5 & 4.9 & 2.0 & 2.8  \\
2 & 2.0 & 6.2 & 2.5 & 3.8		\\
3 & 2.5 & 8.0 & 3.1 &	7.1 \\
4 & 3.0 & 9.5  & 3.9 & 4.7	\\
5 & 4.0 & 12.7  & 5.1 & 5.8	\\
\hline
		\end{tabular}	\label{tab:SummaryOfExperimentalConditions}
\end{table}

The experiments were carried out at constant pressure difference and constant capillary pressure. The experimental setup was shown by Alfazazi~\textit{et al.} \cite{Alfazazi2024}. The total flow rate was measured using a Coriolis-based mass flow meter. Deionized distillated water and decane were injected simultaneously in a fractional flow, $f_w$ = 0.5. A Bronkhorst mass flow meter was used to measure the total flow rate, \textit{i.e.}, the sum of contributions from both phases, at intervals of 30 milliseconds. This means that $\tau_0 = 30 $ millisecond. The input pressure ranged between 1.5 and 4.0 psi (equivalent to 10342 and 27579 Pa, respectively);  the outlet was under ambient conditions. The Capillary number calculated from the average total flux is shown in Table \ref{tab:SummaryOfExperimentalConditions} together with other standard experimental measures.

%=======================================
% Results
%=======================================

\section{ Results}\label{sect:results}

This work presents new procedures for the analysis of volume or flux fluctuation correlations in porous media; see Section \ref{subsect:noiseanalysis}. The flux fluctuations are inputs in the fluctuation-dissipation theorem. The proposed procedure is therefore different from previously published work on pressure fluctuations. Data are obtained for sintered glass beads. These are used to demonstrate the type of physical insight available; see Section \ref{subsect:ganglion}. The familiar power analysis is also presented in Section \ref{subsect:poweranal} to document results. 

\subsection{The multiple resonance memory function}
\label{subsect:noiseanalysis}

\begin{figure*}[ht]
\centerline{\includegraphics[scale=0.4]{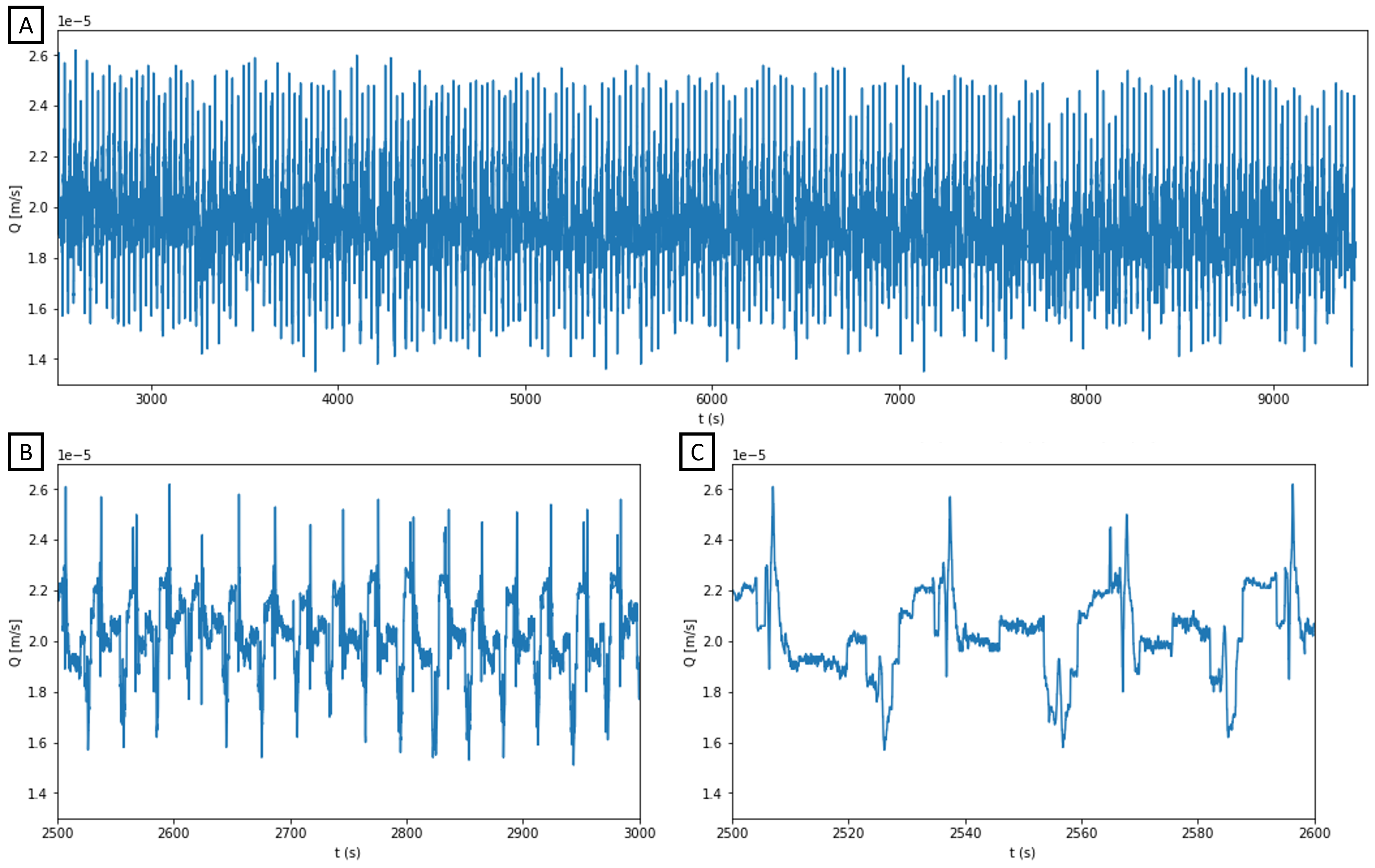}}
	\caption{Time series of the flux $Q$ for the whole time interval (A) and shorter time intervals (B,C) to show more details. Time series is from Experiment 1 reported in Table \ref{tab:SummaryOfExperimentalConditions}}
	\label{fig:fluxtimeseries}
\end{figure*}

Typical time series of total volume flow $Q$ are shown on a second scale in Figure~\ref{fig:fluxtimeseries}A. the relation between $Q$ and $J_V$ was given in Eq.\ref{eq:4.10}. The figure shows the fluctuating flow $Q$ at a constant pressure difference. We see longer and shorter time intervals between the fluctuations. Figure 3A shows fluctuations around an average value of $Q$. The fluctuations are more visible in panels B and C. Regular patterns appear about 25 seconds apart; see panels B and C. 
From the details of Fig. \ref{fig:fluxtimeseries}C, we also see smaller local optima. Evidently, a Fourier transform of the data may be useful, and this motivates the first step in the analysis.
The time series of the volume flux were therefore interpolated to equidistant time intervals, then determined and subtracted from the instantaneous flux value, $Q$ (or $J_V$) to give the noise.
\begin{equation}
   Q_R = Q - <{Q}>
\end{equation}\label{eqn:appaQstar}
where $<{Q}> = 1/N \sum_{i=1}^N Q_i$ is the mean flux for $N$ data points $Q_i$. From  typical data at a pressure drop of 1.5 Psi (experiment number 1), shown in Fig.~\ref{fig:fluxtimeseries}, we find the average value of the flux equal to 2.0 $\times 10^{-5}$ m/s (see Table \ref{tab:SummaryOfExperimentalConditions}). 
The $Q$ is by definition the integral of $J_V$ over the cross section divided by the area of the cross section. Therefore, $Q$ and $Q_R$ will not depend on $y$ and $z$.

The next step was to compute the volume autocorrelation function from the instantaneous values. The volume autocorrelation function is given by
\begin{equation}
C (\tau) = \bigl< Q_R(\tau)Q_R(0) \bigr> = \frac{1}{T}\int_0^{T} Q_R(t+\tau)Q_R(t)dt 
\end{equation}
where $T$ is the period over which the signal is averaged equal to the length of the sample divided by the average volume velocity. This $T$ should not be confused with the temperature. The volume autocorrelation function was calculated from the discrete data set $Q_i$ as 
\begin{equation}
	\bigl< Q_R(\tau)Q_R(0) \bigr> = \frac{1}{N \cdot \sigma_Q}\sum_{i=1}^N Q_{R,i+k}Q_{R,i}
\end{equation}\label{eqn:appaautcorrelationdisc}
where $\tau = k\cdot dt$ with time step $dt$. The function was normalized to the variance $\sigma_Q$ of the data set $Q_{R,i}$, the variance of each experiment is given in Table \ref{tab:SummaryOfExperimentalConditions}. Using Eqs.\ref{eq:34}, and introducing the regular frequency $f = \omega /(2 \pi)$, we conclude that 
\begin{equation}
C(f)=\frac{k_{\text{B}}}{\pi V_L} L_{VV}^{\rm{exp}}(f)  \label{8}
\end{equation}

We now introduce the Lorentzian multiple resonance peak model and obtain the autocorrelation function $C(f)$:
\begin{equation}
C(f )= \sum_{i=1}^{m}
C^{i,0}\left[ \frac{2 f \Gamma^{i}}{ 2 f \Gamma^i - i (( f^i)^2 - f^2)  } \right]   \label{10}
\end{equation}%
The real part is%
\begin{equation}
{Re}C(f )= \sum_{i=1}^{m}
C^{i,0} 
\left[ \frac{ (2 f \Gamma^{i})^2}{ (( f^i)^2 - f^2)^2 + (2 f \Gamma^i)^2 } \right]
\end{equation}%
and the imaginary part%
\begin{equation}
{Im}C(f )= \sum_{i=1}^{m}
C^{i,0} 
\left[ \frac{2 f \Gamma^{i}(( f^i)^2 - f^2)}{ (( f^i)^2 - f^2)^2 + (2 f \Gamma^i)^2 } \right]
\label{12}
\end{equation}
These Lorentzian peak properties can now be used to fit the experimental data. From the fits we obtain for each peak, the resonance frequency $f ^{i}$, the relaxation time $\tau^i$, the bandwidth $\Gamma^i$, where $\tau ^{i} =1/\Gamma^i$. We used real as well as imaginary parts of the resonance frequency to determine the amplitudes $C^{i,0}$:
\begin{equation}
{Re}C(f ^{i})= {C^{i,0}}
\end{equation}
We finally arrive at the expression for the experimentally observed memory function 
\begin{equation}
    L_{VV}^{\rm{exp}}(f) = \frac{\pi V_L}{k_B} C(f) 
    \label{eq:memories}
\end{equation}
This is the frequency dependent permeability. It provides us with the response with an oscillating pressure difference as driving force. 
 We have provided a route to this memory-dependent permeability of Darcy's law with an oscillating pressure gradient. It applies to the porous medium investigated here, as the pre-factor $F= \pi V_L/k_B$ is a function and the geometry of the materials of the REV. In the first study \cite{Alfazazi2024} we reported the zero-frequency permeability  to be a factor $F$ different from the permeability of the linear law. We do not know if the same $F$ applies to all frequencies. This should be investigated. We recall, however, that theory demands that the permeability, also the frequency-dependent permeability is a function of state variables, but not of the driving forces or the fluxes used in the experiments.

\subsection{Properties of the multiple resonance memory-model}
\label{subsect:ganglion}

\begin{figure}[!ht]
	\centerline{\includegraphics[scale=0.5]{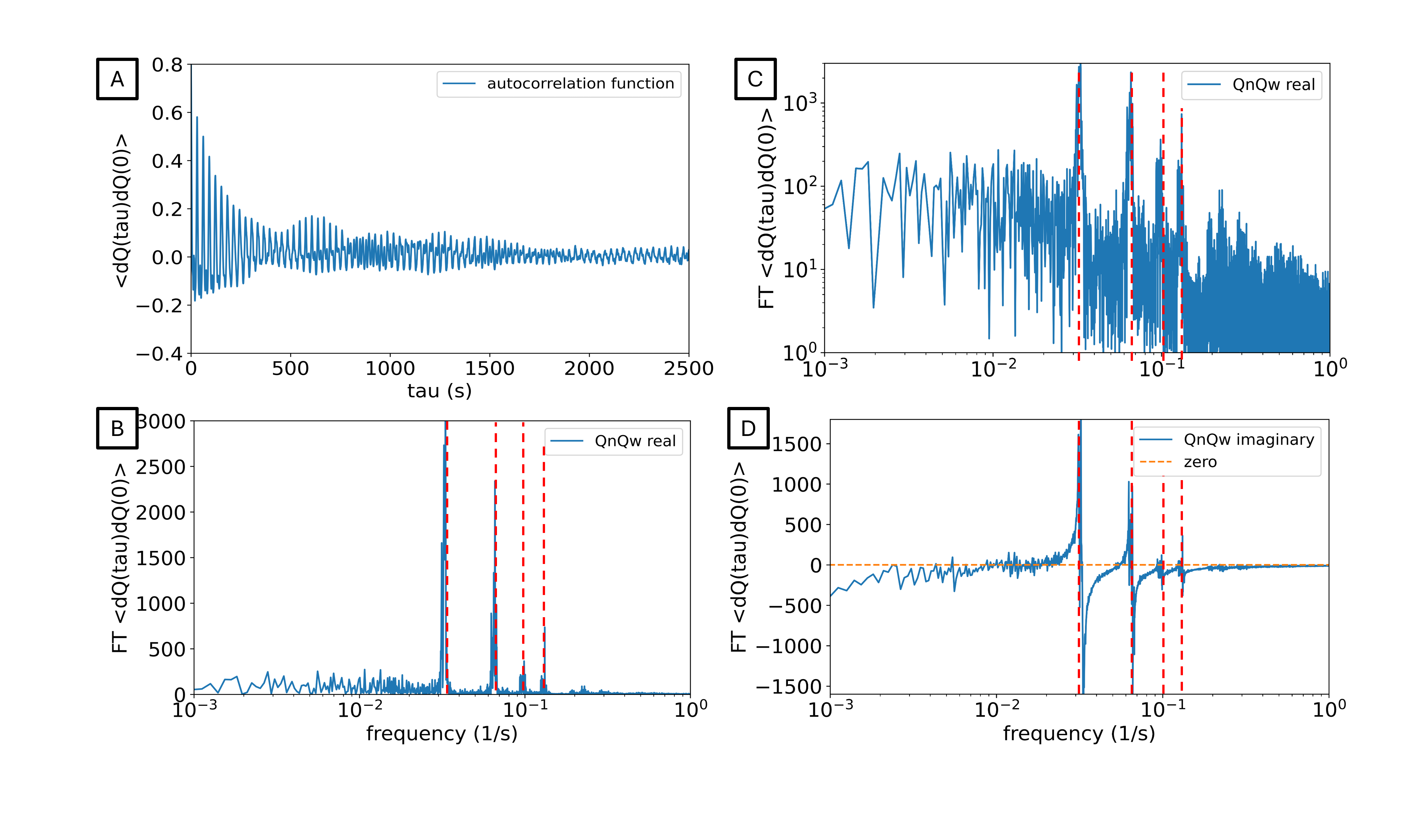}}
	\caption{Auto-correlation function of flux fluctuations as a function of time (A). The real (B,C) and imaginary part (D) of the Fourier transform of the auto-correlation are shown as functions of frequency $f$. The multiphase flow takes place at a pressure drop of 1.5 Psi.}
	\label{fig:FT 1.5 Psi}
\end{figure}

We return to the autocorrelation functions, computed as described above and shown in Figure \ref{fig:FT 1.5 Psi}. We see the oscillating signal with a decay to zero in panel A. Similar behavior was observed for all the experiments reported in Table II. Evidently, the Fourier transform from the time-to-the-frequency domain is in place.

In order to further qualify the findings, Fourier transformations were also used to analyze data from the two-phase flow in the absence of the porous medium. The results are shown in Appendix B for $\Delta P =$1.5 Psi. At the same Ca, the transformed signal showed fewer peaks in the absence of the porous medium than in the presence of the porous medium. Small amplitudes were also observed outside the range reported for the experiments with a porous medium for the same Ca. This suggests that the observed power spectrum is not the result of noise within the Coriolis tube but the actual fluctuations caused by the flow of two immiscible phases through a porous medium. 

Fourier transforms were performed on the data, which transformed the autocorrelation of the total signal into a frequency domain. The results are shown in Fig.\ref{fig:FT 1.5 Psi}B-D. The sum was decomposed into real (B,C) and imaginary parts (D). 
Data reduced in this manner for selected experiments are reported in Table \ref{tab:SummaryOfExperimentalConditions}. Results are shown for the pressure difference of 1.5 psi only. 
The oscillations were evident in Figure \ref{fig:FT 1.5 Psi} and in all other experiments listed in Table \ref{tab:SummaryOfExperimentalConditions}.
We therefore propose that the oscillations arise from similar underlying flow processes which can all be regarded as the signature of the porous medium in use.

It was found that a Fourier transform of the $Q(t)$ time series alone, i.e. without computing the autocorrelation function, does not produce such resonance peaks. The peak shape we observe is mainly defined through the envelope of the time decay in the autocorrelation function, \textit{e.g.} from Fig.~\ref{fig:FT 1.5 Psi}A, while the resonance frequency is defined by the periodic signal therein. For example, the Fourier transform of an exponentially decaying sinusoidal oscillation is a Lorentzian~\cite{Bronstein1981}.
The best fits (using a least-squares fit method in the Python \textit{curve \mathunderscore fit} package) to the experimental data were always obtained with the Lorentzian multiple resonance model described above. The resonance frequencies $f^i$, the bandwidth $\Gamma^i$, and the peak height $C^{i,0}$, all obtained from the fit in the normalized peak area, are listed in Table~\ref{tab:Lorentzianfits} for two pressure drops, 1.5 and 4.0 Psi. The permeability $L_{VV}^{i,0}$ can be found from $C^{i,0}$ and Eq. \ref{eq:memories} once $k_B$ and $V_L$ are known.
Inspecting Table~\ref{tab:Lorentzianfits}, we see from column 7, that the assumption made in the model, $f^{i}\tau ^{i}>>1$, was obeyed everywhere. 

\begin{table}[ht]
\small
\caption{Results from fitting experimental data to the Lorentzian multiple resonance model. The peak frequency $f^i$ and bandwidth $\Gamma^i$ associated with the Lorentzian fits are shown for experiments with pressure drops of 1.5 and 4 PSI (Figs.~\ref{fig:Lorentzian4PSImultipeak}). In the multiple resonance memory function, the relaxation time is $\tau^i$ =$1/\Gamma^i$. } 
	\centering
	\begin{tabular}{c|c|c|c|c|c|c|c} \hline
 $\Delta P$ & Peak  & $f^i$ & $\Gamma^i$ & $\tau^i$ & $C^{i,0}$  & $\tau^i f^i$ & \textgoth{Q} \\
      Psi & -  &  Hz  & Hz & s  &  & - & - \\ 
\hline
1.5	& 1	& 0.033 & 0.00058   &	1716 &	2650 &	56	& 28 \\
1.5	& 2	& 0.066 & 0.00128   &	781	 &   980 &	51	& 26 \\
1.5	& 3	& 0.098 & 0.00344   &	290	&    107 &	28	& 14 \\
%1.5	& 4	& 0.132 & (0.00011) & (9094) & (785) &  (1199)	& (559) \\
\hline							
4.0 &	1 &	0.094 &	0.00106 &	943	& 2235 & 89 &	85 \\
4.0 &	2 &	0.119 &	0.00194	&   515	& 121 &	61	& 31 \\
4.0 &	3 &	0.188 &	0.00196 &	511	& 260 &	96	& 48 \\
4.0 &	4 &	0.282 &	0.00234 &	428	& 31 &	121	& 61 \\
4.0 &	5 &	0.377 &	0.00464 &	216	& 16 &	88	& 44 \\
%4.0 &	6 &	0.480 &	(0.00025) &	(3960) & 9 & (11943) &	(5972) \\
\hline
		\end{tabular}	
  \label{tab:Lorentzianfits}
\end{table}

Return to the real and imaginary parts of the autocorrelation function that are shown in Figure \ref{fig:FT 1.5 Psi}B-D. The real part is shown in panel B and in an enlarged version in panel C, while panel D gives the imaginary part. Consider the resonance peaks of Fig.\ref{fig:FT 1.5 Psi}. Four peaks emerge, well separated, and all within a narrow frequency range (from $10^{-2}$ to 1 Hz). In the real and imaginary parts of Fig. \ref{fig:FT 1.5 Psi} D, we see the same resonance frequencies, indicated by vertical red dashed lines. There is no significant step change after each resonance, as is typical for dielectric phenomena~\cite{Kremer2002}. This finding supports the idea that our model is an analogy to conductance phenomena in general. The model also has an analogy to a set of harmonic oscillators with friction. The behavior predicted from the formulas in Eqs.~\ref{3.19} and \ref{3.20} can be observed in Fig. \ref{fig:FT 1.5 Psi}D and all other experimental results (data not shown). 
This supports the idea of using the resonance peak heights $C^{i,0}$ to find $L_{VV}^{i,0}$ according to Eq.43.

However, some peak groups consist of substructures. For instance, for the flow experiment at $\Delta p=$ 1.5 Psi from Fig.~\ref{fig:Lorentzian4PSI}, the peak group around 0.095 Hz consists of at least 2 subpeaks which can be again fitted= well with the Lorentzian model. However, in general, the emerging picture is that of independent peaks.

The real part is captured well on both a linear and a logarithmic scale. As can be seen in Fig.~\ref{fig:Lorentzian4PSI}, the peak shape is clearly described by a Lorentzian. 
%Note that we used a Lorentzian for the fit in which the peak area was normalized by the factor $\Gamma$~\cite{Berg2002} (CHECK). 

\begin{figure}[!ht]
	\centerline{\includegraphics[scale=0.14]{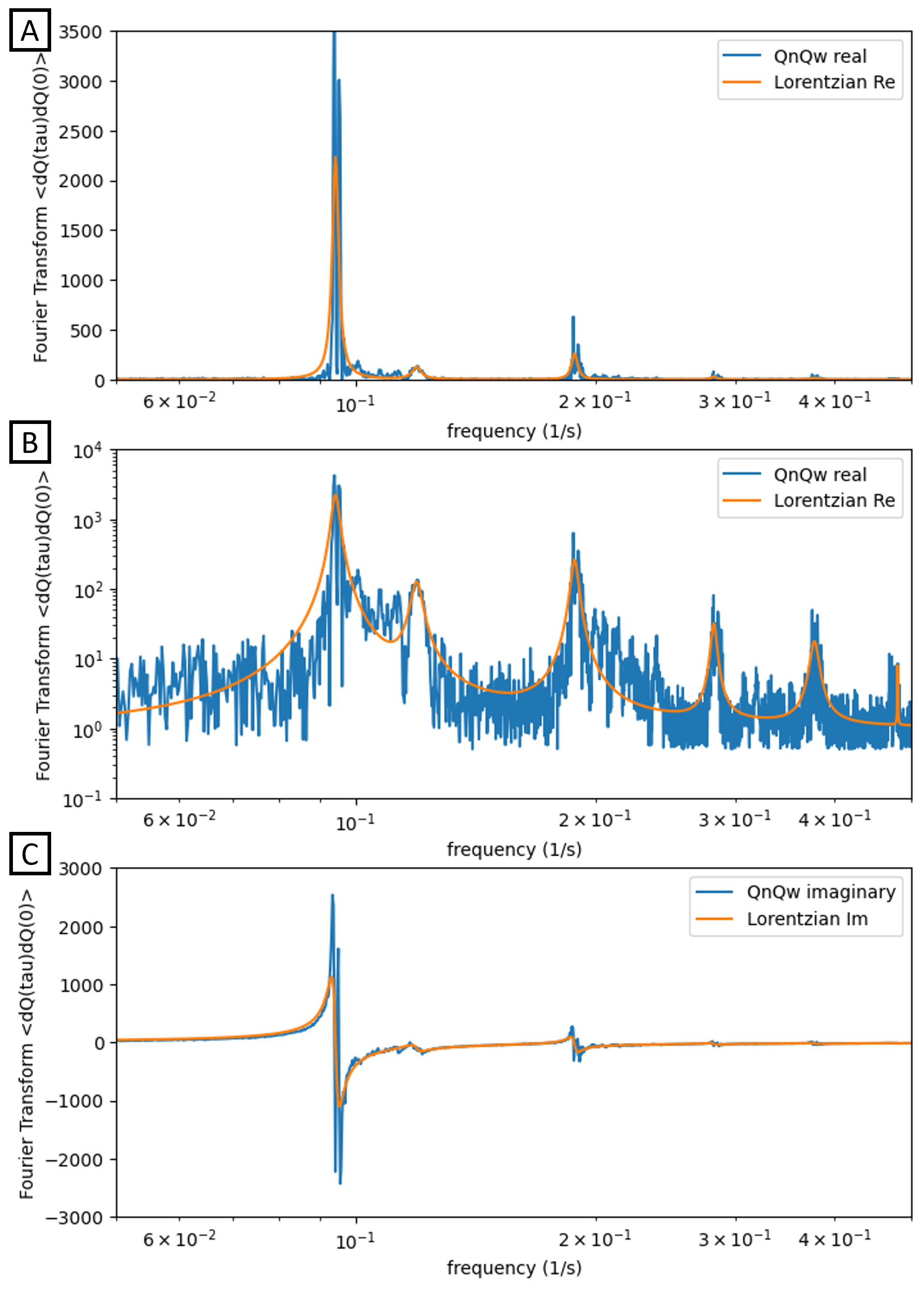}}
	\caption{Real part on a linear (A) and logarithmic scale (B) and imaginary part (C) of the Fourier transform of the autocorrelation function for the multiphase flow experiment at pressure drop of 4.0 Psi and respective fit with a Lorentzian multi-peak model (ignoring the sub-peaks in peak groups). Results are plotted vs frequency $f$.}
	\label{fig:Lorentzian4PSImultipeak}
\end{figure}

\begin{figure*}[!ht]
	\centerline{\includegraphics[scale=0.08]{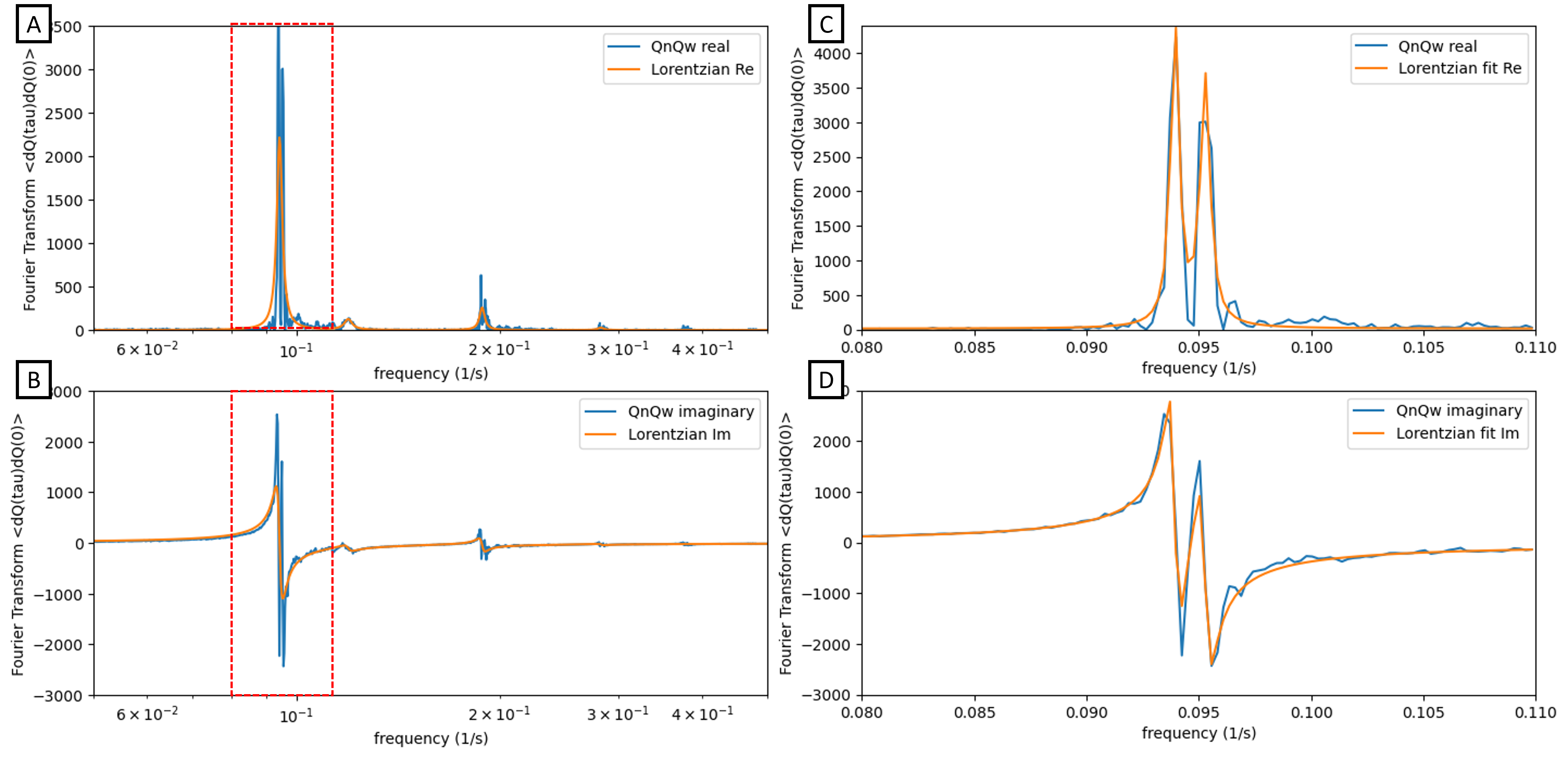}}
	\caption{Real (A) and imaginary part (B) of the Fourier transform of the autocorrelation function for the multiphase flow experiment at a pressure drop of 4.0 Psi, and respective fits with a Lorentzian multi-peak model (ignoring the sub-peaks in peak groups). Results are plotted vs frequency $f$. The individual sub-peaks in the peak group around 0.095 Hz (red frame in C,D) are fitted with the multiple resonance model using a sum of 2 Lorentzians. }
	\label{fig:Lorentzian4PSI}
\end{figure*}

A comment on the actual numbers in Table \ref{tab:Lorentzianfits} remains.
As explained before, the shape of the peak (Lorentzian) and hence also the bandwidth are associated with the envelope of the flux autocorrelation function. As can be seen in Fig.~\ref{fig:FT 1.5 Psi}A the envelope shows an overall decay over time, meaning that there is a finite correlation time, which is then represented by the bandwidth $\Gamma^i$ of the respective Lorentzian peak. This finite correlation time in the autocorrelation function is then interpreted as a relaxation time of the underlying resonator. 
The relaxation times are with a few exceptions below 1 hr (3600 s), describing phenomena relaxing on the minute scale. The bandwidth $\Gamma^i$ of the dominant peaks is between 0.0006 and 0.0046 Hz, which implies that the associated relaxation time \cite{Berg2003}
\begin{equation}
    \tau^i = \frac{1}{\Gamma^i}
\label{eqn:bandwidthrelaxationtime}
\end{equation}
\noindent is in the range of 4-29 minutes. This fits very well with ganglion dynamics (or intermittency), as will be discussed in the following.  Amplitudes of this kind may be subject for further systematic research, e.g. on their dependence of the porous media properties. 

When trying to associate observed resonances with underlying physical phenomena, the quality factor $\textgoth{Q}$  of the resonances may be helpful:
\begin{equation}
    \textgoth{Q} = \frac{f^i}{2\Gamma^i}
\end{equation}\label{eqn:qualityfactor}

For our experiments $\textgoth{Q} \approx 15-60$ (see Table \ref{tab:Lorentzianfits}). For comparison, quartz crystal microbalances have $\textgoth{Q}$-factors between 1E3 and 1E6, the latter in vacuum, and an acoustic tuning fork has a quality factor around 1000.

\subsection{Power spectral analysis}
\label{subsect:poweranal}

Power spectral analysis can also be used to obtain insight into the different transport processes that take place within the porous media. Here, in Figures \ref{fig:FT 1.5 Psi}, we observe the distribution of power between different frequencies and time scales. Table \ref{tab:FT para} summarizes the results for the power distributions for all the experiments listed in Table \ref{tab:SummaryOfExperimentalConditions}. Table \ref{tab:FT para} highlights the observed resonance peaks, their respective frequency range, the associated signal-to-noise ratio, and the relative power of the dominant frequencies. Relative power was determined by calculating the total power of the entire power spectrum for each given experiment (for example, Fig.\ref{fig:FT 1.5 Psi}B). The relative power is then calculated for specific frequency ranges (with the most pronounced resonances). This is done by dividing the total power within the specified frequency range by the total power of the entire spectrum. Similarly, the signal-to-noise ratio (expressed in decibels dB) is computed as the ratio of the signal power within each specified frequency range to the noise power, which is estimated as the average power of the other values within that range but excluding the signal power.  

\begin{table*}[ht]
\footnotesize
\centering
\caption{Summary of Power Distribution for Different Experiments in the Presence of Porous Media}
\begin{tabular}{cccccc}
\hline
\begin{tabular}[c]{@{}c@{}}Exp\\No\end{tabular} & \begin{tabular}[c]{@{}c@{}}Peak \\Amplitude\end{tabular} & \begin{tabular}[c]{@{}c@{}}Corresponding\\Frequency Range\end{tabular} & \begin{tabular}[c]{@{}c@{}}Relative \\Power (\%)\end{tabular} & \begin{tabular}[c]{@{}c@{}}Signal to Noise \\Ratio (SNR)\end{tabular} & \begin{tabular}[c]{@{}c@{}}Relation to\\Flow Dynamics\end{tabular} \\
 \\ \hline \hline
 
1  & \begin{tabular}[c]{@{}c@{}}above 3000\\2000\\500\\below 200\end{tabular} & \begin{tabular}[c]{@{}c@{}}0.03 - 0.04\\0.06 - 0.08\\0.09 - 0.10\\0.10 - 0.10\end{tabular} & \begin{tabular}[c]{@{}c@{}}56.92\\26.01\\1.65\\5.24\end{tabular}  & \begin{tabular}[c]{@{}c@{}}10.70\\10.96\\10.28\\18.73\end{tabular}  & \begin{tabular}[c]{@{}c@{}} \\ganglion dynamics\\collective phenomena\\ \end{tabular}\\
\hline

2  & \begin{tabular}[c]{@{}c@{}}3000\\2000\\300\\below 100\end{tabular} & \begin{tabular}[c]{@{}c@{}}0.03 - 0.04\\0.05 - 0.06\\0.09 - 0.10\\0.10 - 0.20\end{tabular} & \begin{tabular}[c]{@{}c@{}}2.10\\88.51\\0.19\\3.54\end{tabular}   & \begin{tabular}[c]{@{}c@{}}11.59\\13.45\\6.51\\18.08\end{tabular}  & \begin{tabular}[c]{@{}c@{}} \\ganglion dynamics\\collective phenomena\\ \end{tabular}\\
\hline

3  & \begin{tabular}[c]{@{}c@{}}2500\\1000\\300\\below 100\end{tabular} & \begin{tabular}[c]{@{}c@{}}0.05 - 0.06\\0.08 - 0.10\\0.10 - 0.20\\0.20 - 0.50\end{tabular} & \begin{tabular}[c]{@{}c@{}}1.23\\79.27\\11.11\\3.36\end{tabular}  & \begin{tabular}[c]{@{}c@{}}8.63\\17.50\\18.35\\9.50\end{tabular}  & \begin{tabular}[c]{@{}c@{}} \\ganglion dynamics\\collective phenomena\\ \end{tabular} \\
\hline

4  & \begin{tabular}[c]{@{}c@{}}2500\\below 1000\\below 300\end{tabular} & \begin{tabular}[c]{@{}c@{}}0.06 - 0.09\\0.10 - 0.20\\0.20 - 0.30\\0.30 - 0.40\end{tabular} & \begin{tabular}[c]{@{}c@{}}88.50\\7.21\\1.33\\0.57\end{tabular}   & \begin{tabular}[c]{@{}c@{}}13.67\\18.93\\12.69\\5.13\end{tabular} & \begin{tabular}[c]{@{}c@{}} \\ganglion dynamics\\collective phenomena\\ \end{tabular} \\
\hline

5  & \begin{tabular}[c]{@{}c@{}}below 1000\\below 200\end{tabular} & \begin{tabular}[c]{@{}c@{}}0.06 - 0.08\\0.10 - 0.20\\0.20 - 0.30\\0.30 - 0.50\end{tabular} & \begin{tabular}[c]{@{}c@{}}1.0\\50.87\\25.34\\11.47\end{tabular}   & \begin{tabular}[c]{@{}c@{}}7.95\\8.24\\11.15\\11.18\end{tabular} & \begin{tabular}[c]{@{}c@{}} \\ganglion dynamics\\collective phenomena\\ \end{tabular} \\
\hline
\end{tabular}
\label{tab:FT para}
\end{table*}

From Table \ref{tab:FT para}, when the pressure drop is 1.5 Psi (Ca = $4.9 \times 10^{-6}$), the dominant peaks are observed in the range of 10$^{-2}$ to $10^0$ Hz. The highest amplitudes (largest peaks) are observed between 0.03 and 0.1 Hz. The other experiments display similar behavior with a slight reduction in peak amplitude and a slight increase in the corresponding frequency ranges as the capillary number increases. Table \ref{tab:FT para} also gives the relative power of each range in percent along with the signal-to-noise ratio for each experiment. Relative power results suggest that the first frequency range is likely more important than the subsequent ones. However, the signal-to-noise ratio suggests that all identified peaks are distinguishable from the background noise.

The frequencies observed here are roughly in the range of $1E-2$ and $1E-1$ Hz, placing the underlying phenomena according to Fig.~\ref{fig:frequencyoverview} in the range of ganglion dynamics and other collective and repetitive phenomena \cite{Spurin2023DMD}, sometimes referred to as intermittency~\cite{Spurin2021}, but potentially also traveling wave solutions \cite{Rucker2021}. The associated underlying flow dynamics are therefore suggested in the final column of Table \ref{tab:FT para}.

%=======================================

\section{Discussion}\label{sect:discussion}

 We have presented a new procedure for the analysis of flux fluctuations. The Darcy law permeability with its frequency dependence was already obtained for zero frequency by\cite{Alfazazi2024}. Here we report the associated frequency-dependent memory function, which can be applied to model flow with oscillating driving forces. 
 
 The work was inspired by the network simulations of Winkler \textit{et al} \cite{Winkler2020}. Winkler \textit{et al.} simulated self-correlation and cross-correlation coefficients in a honeycomb lattice with two-phase flow and found a symmetric matrix of transport coefficients. The pore flow was modeled with the Washburn equation and the flux-flux correlation functions were obtained on a time scale around 10$^{-3}$s. 
The present theoretical description applies to the regime of linear flux-force relations and uses the bold assumption of local equilibrium in the REV. Outside of this regime, conditions for validity must be further specified. However, once the theory has proven useful as a tool to obtain permeabilities, we may also reverse the argument and state that the bold assumption about the REV may be justified. This claim may have an impact outside the present field of analysis.

Because the FDT refers to flux-flux correlations, our experiment was performed with a constant pressure difference, while most experiments reported in the literature have not been performed under this boundary condition~\cite{Schluter2017,Armstrong2014,spurin2023pore}. Therefore, a direct comparison is difficult. In general, we have worked under the hypothesis that the values of the coefficients obtained reflect properties of the underlying flow mechanism, and not some unrelated noise. This should also be true for the memory function reported here. Typical mechanisms for porous media flow include phenomena such as Haines jumps and traveling waves; see Fig.~\ref{fig:frequencyoverview}. In the present case (a sintered glass bead pack with decane and water), the data extracted from our theory are pointing to underlying flow mechanisms on a time scale of minutes.

In our further analysis of the memory function, we chose a multiple-peak Lorentzian model. On this background we can now ask; How can phenomena that are individually not in phase and with individual relaxation times of the order of 1-2 s~\cite{Armstrong2014} produce such sharp resonances with quality factors $\textgoth{Q} \approx 15-60$ (see Table \ref{tab:Lorentzianfits}). The quality factor is the ratio of the time scale of the periodic oscillations in the time signal to the time scale over which these oscillations lose their coherence. After some time, oscillations are out of phase with oscillations at the beginning of the time series, meaning they become uncorrelated. This is the decay time in the autocorrelation function in Fig.~\ref{fig:FT 1.5 Psi}A. Resonances with similar quality factor have been associated with traveling saturation waves~\cite{Rucker2021} when pressure transducer measurements are analyzed. 

Traveling waves~\cite{Rucker2021} caused by a Darcy scale 1D instability~~\cite{Plohr2001, Corli2018, Mitra2020} could be a mechanism that causes oscillations and provides phase coherence between individual ganglion dynamics events, since the energy scale of the traveling wave associated with the saturation change due to displacement is sufficient to trigger Haines jumps (cascading)~\cite{Rucker2021}. The next question is related to the peak form. This is very well fitted to a Lorentzian form, similar to the peaks of quartz crystal resonators~\cite{Berg2002}. In general, our Lorentzian peak agreed well with a high-quality / weakly damped resonator with large values of $\textgoth{Q}$. 

However, a number of open discussions remain with respect to the model and a possible analogy with other relaxation phenomena~\cite{Kremer2002}. We are looking for a model that may help establish the concept of memory in an intuitive manner. The closest analogy would be the dielectric function $\hat{\epsilon}$ in which the real part consists of steps between resonances on the frequency scale. This reflects the storage of charge and energy; and a storage term is consistent with memory effects. 

The functional form used to fit the data in this work is commonly used in quartz crystal resonators~\cite{Berg2002}, but is better represented by an analogue of electrical conductivity (conductivity
$\hat{\sigma}=-i\omega (\hat{\epsilon}-1)/4\pi$). However, this analog does not contain steps and small steps may be present in our data; see panel C in Fig.~\ref{fig:FT 1.5 Psi}.
The steps are now not captured by a fit to the multi-resonance model (the conductivity-based model).
In addition, even though the general peak form is Lorentzian, the normal peak form of the real part is a ''dispersion'' curve, and the imaginary part is the actual peak. 
Our data, on the other hand, have a peak for the real part and a dispersion curve for the imaginary part. 
This behavior is reflected in the autocorrelation analysis of time series, e.g., also for a synthetic data set consisting of a sinusoidal oscillation. 
In summary, the occurrence of steps in the real part is not fully explained by a conductivity-based multiresonance model. 

In general, more work will be needed to fully explain the observed phenomena and to further narrow the link to the underlying flow phenomena. The typical length and time scales for multiphase flow in porous media are a challenge. 

\section{Conclusions}\label{sect:conclusions}
In this work, we provide the first experimental support for a newly developed theory for two-phase flow in porous media with memory. A sound theoretical basis (or framework) has been provided in non-equilibrium thermodynamic theory for the analysis of volume autocorrelations inherent in fluctuation dissipation theorems of the transport process.  We have given a first example on how we can extract information on the transport properties of multiphase flow in porous media from such flux fluctuations. 

For the two-phase flow of water and decane in a matrix of glass beads with particle sizes ranging from 0.09 to 0.15 mm, we find that the multiple peak resonance function for the permeability is of a Lorentzian type, with relaxation times on the order of minutes to hours. This points to collective phenomena as the underlying mechanism, which is plausible for the experimental system in use. The system is able to store or release energy in terms of surface area variations and dissipate energy in the process, as one may expect from highly immiscible components such as decane and water.

The description applies to the Darcy level, unlike other bottom-up upscaling approaches. Green-Kubo-like formulas are obtained and shown to work. Theoretical derivations are based on the assumption of local equilibrium in a representative elementary volume (REV). The REV variables are coarse-grained variables on the Darcy scale and obey the Gibbs equation \cite{Kjelstrup2019,Bedeaux2022}. 
This assumption, here adapted to porous media, is standard in nonequilibrium thermodynamics and enables us to derive the entropy production and the constitutive equations for the REV\cite{deGroot1984,Kjelstrup2020,Bedeaux2022}. Fluctuation-dissipation theorems were formulated accordingly, for the first time, for porous media with memory. 
Elsewhere \cite{Bedeaux2022,Bedeaux2023}, we have shown that the integrals of the correlation functions resulted in $2k_{\rm{B}}$ times the Onsager conductivities.  The $L_{VV}$ coefficients used in the present work are expected to do the same. They refer to a particular REV and are functions of the state variables of this REV. 

Time-resolved data on flux-flux correlations, even on pressure fluctuations ~\cite{Rucker2021} may provide a more detailed insight into the substructure, of multiphase flow in porous media, and how it varies with the physical-chemical properties of the pores and fluids. These and associated underlying capillary phenomena have in the past been dismissed as noise or associated with pore-scale phenomena that would not affect Darcy-scale transport~\cite{Rucker2021}. 

This study has shown that there is more to gain by applying the fluctuation dissipation theorems.
However, more systematic applications of the present theory would be instructive. Both network simulations and experimental studies would aid in the development of the theory developed herein. Our current experimental results provide only the first support and motivation for the proposed theoretical development. Using simple experiments, we have demonstrated that Green-Kubo-like expressions can be formulated for a two-phase fluid flow driven by a constant pressure drop. The resulting autocorrelation functions demonstrated memory effects, and further analysis of the power spectrum demonstrated Lorentzian behavior. Here, the observed frequencies suggested ganglion dynamics and/or other collective behavior, sometimes called intermittency \cite{Spurin2021}. This suggests that fractional-flow systems have different modes of transport. The FDT may also provide a means to explain the impact of these modes on the actual permeability of various other media. Future research on this topic is expected to provide rich insights.

\section*{Acknowledgments}\label{sect:acknowledgements}
The authors are grateful to the Research Council of Norway for their Center
of Excellence Funding Scheme, project no 262644 PoreLab. R.T.A. acknowledges his Australian Research Council Future Fellowship (FT210100165).

%=======================================
% Appendix
%=======================================
\appendix

\section{Entropy production in a porous medium REV}
\label{sect:entropyproduction}

The following text gives a short repetition of the governing equations for a REV scale description of transport in porous media used in this work
\cite{Kjelstrup2018,Kjelstrup2019,Bedeaux2022,Bedeaux2023}. The extensive variables used are coarse-grained variables that describe the REV on the macroscale \cite{Kjelstrup2018,Bedeaux2023}. The terminology that follows these works is common in the field of non-equilibrium thermodynamics (NET)  \cite{Kjelstrup2019,Kjelstrup2020}. 

The representative elementary volume (REV) is considered to be in local equilibrium,
in the sense that the variables (temperature $T$, volume $V^{\rm{REV}}$,
chemical potentials $\mu _{i}$, etc.) obey the Gibbs equation. The REV is
chosen to be large enough to be macroscopic, but not unnecessarily large. In terms
of densities for the porous medium with $n$ independent components, the Gibbs equation is equal to:%
\begin{equation}
\frac{\partial s}{\partial t}=\frac{1}{T}\frac{\partial u}{\partial t}-\frac{%
1}{T}\sum_{i=1}^{n}\mu _{i}\frac{\partial c_{i}}{\partial t}  \label{1.1}
\end{equation}%
Gradients in energy- and molar densities produce changes in the variables on
the macro-scale. These lead to the transport of heat and mass. On the macroscale, we may have a variation in pressure in addition to the gravitational field. A gradient in composition may also be present. In the fractured carbonaceous Ekofisk oil field, a geothermal gradient was also found to be relevant \cite{Holt1983}. The interplay of several driving forces can be central. Here, we are concerned with the equations that govern only mass transport. The analysis follows the standard procedure in NET.  Confinement effects can be included \cite{Bedeaux2023}.

The balance equations for the components and the internal energy of a REV are 
\begin{eqnarray}
\frac{\partial c_{i}}{\partial t} &=&-\frac{\partial }{\partial x}J_{i} 
\nonumber \\
\frac{\partial u}{\partial t} &=&-\frac{\partial }{\partial x}J_{u}=-\frac{%
\partial }{\partial x}\left[ J_{q}^{^{\prime }}+\sum_{i=1}^{n}J_{i}H_{i}%
\right]  \label{1.2}
\end{eqnarray}%
The transport on the REV-scale is here in the $x-$direction only. The molar
fluxes, $J_{i}$ in mol.m$^{-2}$s$^{-1}$, and the flux of internal energy, $J_{u}$ in J.m$^{-2}$.s$^{-1}$, are all macroscale fluxes. The internal energy flux is the sum of the measurable
(or sensible) heat flux, $J_{q}^{^{\prime }}$ and the partial specific
enthalpies (latent heat), $H_{i}$ (in J.mol$^{-1}$) times the molar fluxes, $%
J_{i}$, see \cite{deGroot1984,Kjelstrup2020,Gray1998} for further explanations. The component $r$ (the porous
medium) does not move and is used as a reference frame for the fluxes.
Therefore, the sum in Eq.\ref{1.2} b is only over the fluid phases. The
balance equations for the component and the internal energy densities have
their usual form.

The REV scale balance equation for the entropy is:%
\begin{equation}
\frac{\partial s}{\partial t}=-\frac{\partial }{\partial x}J_{s}+\sigma
\label{1.3}
\end{equation}%
Here $J_{s}$ is the entropy flux in J.K$^{-1}$.m$^{-2}$.s$^{-1}$ and $\sigma $ is the entropy production in J.K$^{-1}$.m$^{-3}$.s$^{-1}$.
The entropy production is positive definite, $\sigma \geq 0$ (the second law of thermodynamics).

We can now proceed to derive expressions for $\sigma $ in the standard way
\cite{deGroot1984,Kjelstrup2020}, by combining the balance equations with the Gibbs equation. We introduce
the balance equations for components and energy in the Gibbs equation; see
\cite{deGroot1984,Kjelstrup2020} for details. By comparing the result with the entropy balance, Eq.\ref%
{1.3}, we identify \cite{Kjelstrup2020} the entropy flux 
\begin{equation}
J_{s}=\frac{J_{u}-\sum_{i=1}^{n}\mu _{i}J_{i}}{T}=\frac{1}{T}J_{q}^{\prime
}+\sum_{i=1}^{n}J_{i}S_{i}  \label{1.4}
\end{equation}%
and the entropy production. 

In systems with a constant temperature as here we find: 
\begin{equation}
\sigma =-\frac{1}{T}\sum_{i=1}^{n}J_{i}\frac{\partial }{\partial x}\mu _{i}
\label{1.5}
\end{equation}

An important observation for systems with memory is that all the
thermodynamic fluxes and forces in the entropy production are time
dependent. In the systems considered, the distribution of the solid phase of the porous medium
is homogeneous. This implies $\partial \mu _{r}/\partial x=0$.

\section{Flow behavior of Coriolis flow meter in the absence of porous media}

\setcounter{figure}{0}
\renewcommand{\thefigure}{B.\arabic{figure}}

In the following, we provide experimental data for the experimental setup without the porous media. The experiments were conducted to establish baseline measurements for when two phases pass through the Coriolis flow meter. Both fluids meet at a T-junction, followed by a single line to the Coriolis flow meter. The fluids were injected at flow rates similar to those of the experimental studies conducted with the porous core. In this paper, we show a typical example with a pressure drop of 1.5 Psi in Fig. B.1. Table B.4 is a summary of the observed resonance peaks, their respective frequency range, the relative power of the dominant frequencies, and the associated signal-to-noise ratio for different pressure drops without porous media. Evidently, the dominant frequencies where the power spectra are observed are not in the same range for experiments with the porous media. 

\begin{figure}
	\centerline{\includegraphics[scale=0.5]{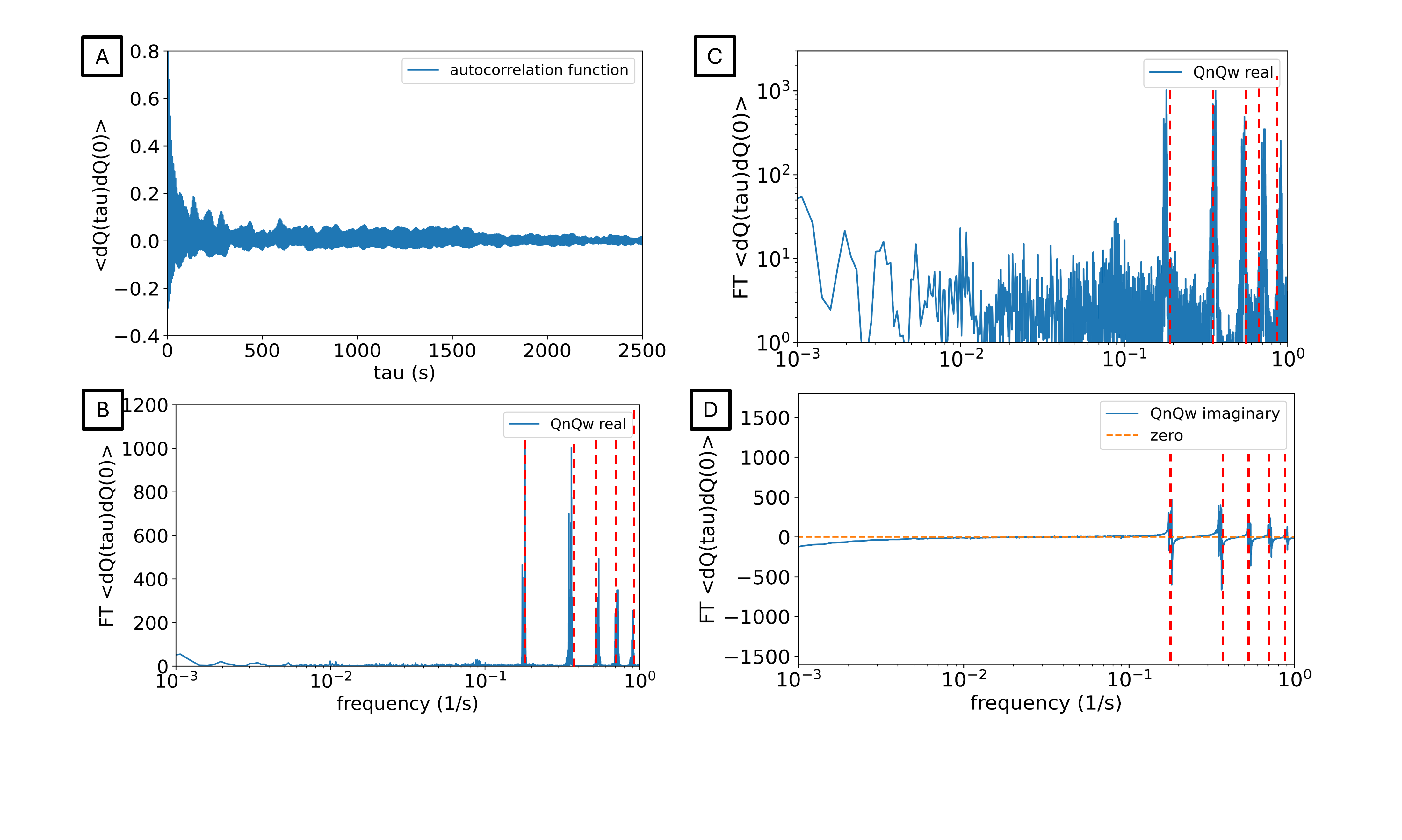}}
	\caption{Auto-correlation functions of flux fluctuations as a function of time (A) and Fourier transform of real (B,C) and imaginary part (C) for multiphase flow at a pressure drop of 1.5 Psi in the absence of a porous media. Results are plotted vs frequency $f$.}
\label{fig:APP 1} 
\end{figure}

\begin{table*}[ht]
\centering
\caption{Summary of Power distribution at Different Frequencies in the Absence of Porous Media}
\begin{tabular}{ccccc}
\hline
\begin{tabular}[c]{@{}c@{}}Pressure\\Drop (Psi)\end{tabular} & \begin{tabular}[c]{@{}c@{}}Peak \\Amplitude\end{tabular} & \begin{tabular}[c]{@{}c@{}}Corresponding\\Frequency Range\end{tabular} & \begin{tabular}[c]{@{}c@{}}Relative \\Power (\%)\end{tabular} & \begin{tabular}[c]{@{}c@{}}Signal to Noise \\Ratio (SNR)\end{tabular} 
 \\ \hline \hline
 
1.5  & \begin{tabular}[c]{@{}c@{}}1000\\500\\200\\below 100\end{tabular} & \begin{tabular}[c]{@{}c@{}}0.1 - 0.2\\0.2 - 0.3\\0.3 - 0.4\\0.4 - 0.5\end{tabular} & \begin{tabular}[c]{@{}c@{}}20.51\\11.94\\45.67\\15.17\end{tabular}  & \begin{tabular}[c]{@{}c@{}}16.33\\15.59\\16.18\\14.37\end{tabular}\\
\hline

2.0  & \begin{tabular}[c]{@{}c@{}}below 1000\\below 200\end{tabular} & \begin{tabular}[c]{@{}c@{}}0.2 - 0.3\\0.4 - 0.5\\0.6 - 0.7\\0.7 - 0.9\end{tabular} & \begin{tabular}[c]{@{}c@{}}30.57\\40.68\\8.88\\11.08\end{tabular}   & \begin{tabular}[c]{@{}c@{}}15.40\\15.93\\11.17\\8.64\end{tabular}\\
\hline

5.0  & \begin{tabular}[c]{@{}c@{}}below 1000\end{tabular} & \begin{tabular}[c]{@{}c@{}}0.5 - 0.6\\0.9 - 1.1\\1.4 - 2.0\\2.0 - 3.0\end{tabular} & \begin{tabular}[c]{@{}c@{}}3.29\\23.64\\40.27\\26.38\end{tabular}  & \begin{tabular}[c]{@{}c@{}}16.12\\13.43\\17.05\\16.34\end{tabular}\\
\hline

8.0  & \begin{tabular}[c]{@{}c@{}}below 500\end{tabular} & \begin{tabular}[c]{@{}c@{}}0.5 - 0.7\\1.0 - 1.1\\1.2 - 2.0\\2.0 - 4.0\end{tabular} & \begin{tabular}[c]{@{}c@{}}60.60\\1.89\\11.55\\18.17\end{tabular}   & \begin{tabular}[c]{@{}c@{}}16.09\\16.12\\17.45\\15.10\end{tabular} \\
\hline
\end{tabular}
\label{tab:APP B1}
\end{table*}


\cleardoublepage
\newpage

 \bibliographystyle{elsarticle-num} 
 \bibliography{FDMemoryExtended}

\end{document}